\documentclass[preprint,sort&compress]{article}
\usepackage[left=2.0cm,right=2.0cm,top=1cm,bottom=2cm,bindingoffset=0cm]{geometry}
\usepackage{float,eepic,subfigure,subfig}
\usepackage{graphicx, eepic}
\usepackage{amsmath,amssymb,amsthm}
\usepackage[hyphens]{url}
\usepackage{listings}
\usepackage{morefloats,placeins}

\begin{document}

\title{Hidden and self-excited attractors in Chua circuit:
\\ SPICE simulation and synchronization}

\author{
M.A. Kiseleva, E.V. Kudryashova,
N.V. Kuznetsov\thanks{Corresponding author. Email: nkuznetsov239 at gmail.com},
O.A. Kuznetsova, \\
G.A.~Leonov,
M.V. Yuldashev,
R.V. Yuldashev
}
\maketitle

\begin{abstract}
Nowadays various chaotic secure communication systems based on
synchronization of chaotic circuits
are widely studied.
To  achieve synchronization, the control signal
proportional to the difference between the circuits signals,
adjust the state of one circuit.
In this paper the synchronization of two Chua circuits
is simulated in SPICE.
It is shown that the choice of control signal is be not straightforward,
especially in the case of multistability and hidden attractors.

\end{abstract}

\section{Introduction: hidden and self-excited attractors in Chua circuit}
Since the first chaotic behavior in dynamical systems was  revealed
by numerical integration \cite{Lorenz-1963},
the researchers started to be interested in circuit implementation of chaos
(see, e.g. \cite{Tokunaga-1989,Robinson-1990} and others),
which allows one to ensure that the pseudo-orbits can be traceable by actual
orbits (see, e.g. the corresponding discussion on shadowing in
\cite{Pilyugin-2011,Lozi-2013}).
At the same time various engineering perspectives of chaotic
circuits application have been found \cite{ChenU-2002}.

\begin{figure}[H]
  \centering
  \includegraphics[scale=0.6]{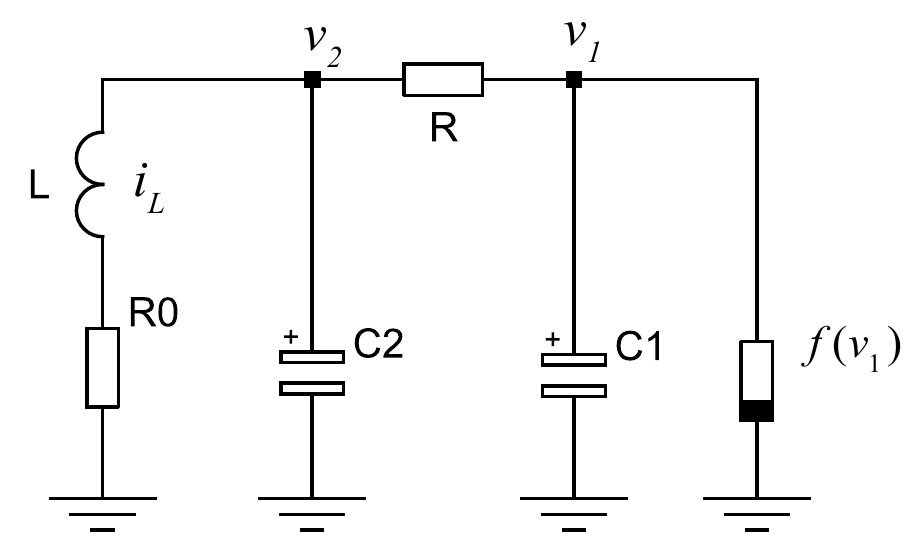}
  \caption{Chua circuit}
  \label{chua-circuit-pic}
\end{figure}

The Chua circuit, invented in 1983 by Leon Chua \cite{Matsumoto-1984,Chua-1992},
is the simplest electronic circuit exhibiting chaos.
Consider one of classical Chua circuits shown in Fig.~\ref{chua-circuit-pic}.
The circuit consists of passive resistors ($R$ and $R_0$),
capacitors ($C_1$ and $C_2$), conductor $L$,
and one  nonlinear element with characteristics $f(\cdot)$,
called Chua diode.
It is described by the following equations
\begin{equation}
\begin{aligned}
\label{chua-circuit-equations}
& \frac{dv_1}{dt} = \frac{1}{C_1} \left(\frac{1}{R} (v_2 - v_1) - f(v_1)\right),
\\
& \frac{dv_2}{dt} = \frac{1}{C_2} \left(\frac{1}{R} (v_1 - v_2) + i_L\right),
\\
& \frac{di_L}{dt} = - \frac{1}{L} (v_2 + R_0 i_L),
\\
& f(v_1) = \left\{
              \begin{array}{lll}
                 G_b v_1 + (G_b - G_a) E_1,  & \text{if} & v_1 \leq -E_1, \\
                 G_a v_1,                    & \text{if} & |v_1| <  E_1, \\
                 G_b v_1 + (G_a - G_b) E_1,  & \text{if} & v_1 \geq E_1. \\
              \end{array}\right.
\end{aligned}
\end{equation}
Here $v_1$ and $v_2$ are voltages across capacitors $C_1$ and $C_2$, respectively,
$i_L$ is a current through conductor $L$,
function $f(v_1)$ is volt-ampere characteristics of Chua's diode.
In the following discussion we choose $C_2$ such that $R C_2 = 1$
and put $E_1 = 1$. By the introduction of new variables
\begin{equation}
\begin{aligned}
& x = v_1, \ y = v_2, \ z =  R i_L,
\\
& \alpha = \frac{1}{RC_1}, \beta = \frac{R}{L}, \gamma = \frac{R_0}{L}
\\
& \psi(x) = R f(x), \ R G_a = m_0, \ R G_b = m_1,
\end{aligned}
\end{equation}
system \eqref{chua-circuit-equations} is transformed to the following form
\begin{equation}
\label{chua-dimensionless}
\begin{aligned}
& \frac{dx}{dt} = \alpha(y - x - \psi(x)),
\\
& \frac{dy}{dt} =  x - y + z,
\\
& \frac{dz}{dt} = - (\beta y + \gamma z),
\\
& \psi(x) = m_1 x + (m_0 - m_1)\text{sat}(x),
\\
& \text{sat}(x) = \left\{
              \begin{array}{lll}
                 -1,    & \text{if} & x \leq -1, \\
                 x,     & \text{if} & |x| <  1, \\
                 1,   & \text{if} & x \geq 1. \\
              \end{array}\right.
\end{aligned}
\end{equation}

Until recently there had been found Chua attractors, which
are excited from unstable equilibria only and, thus,
can be easily computed
(see, e.g a gallery of Chua attractors in \cite{BilottaP-2008}).
Note that L.~Chua \cite{Chua-1992}, analyzing various cases
of attractors in Chua's circuit,
did not admit the existence of attractors of another type
--- so called \emph{hidden attractors},
being discovered later in his circuits.
An attractor is called a \emph{self-excited attractor}
if its basin of attraction
intersects an arbitrarily small open neighborhood of equilibrium,
otherwise it is called a \emph{hidden attractor}
\cite{LeonovKV-2011-PLA,LeonovK-2013-IJBC,LeonovKM-2015-EPJST,Kuznetsov-2016}.
Hidden attractor has basin of attraction
which does not overlap with an arbitrarily small vicinity of equilibria.

\begin{figure}[ht]
  \center
  \includegraphics[width=0.45\textwidth]{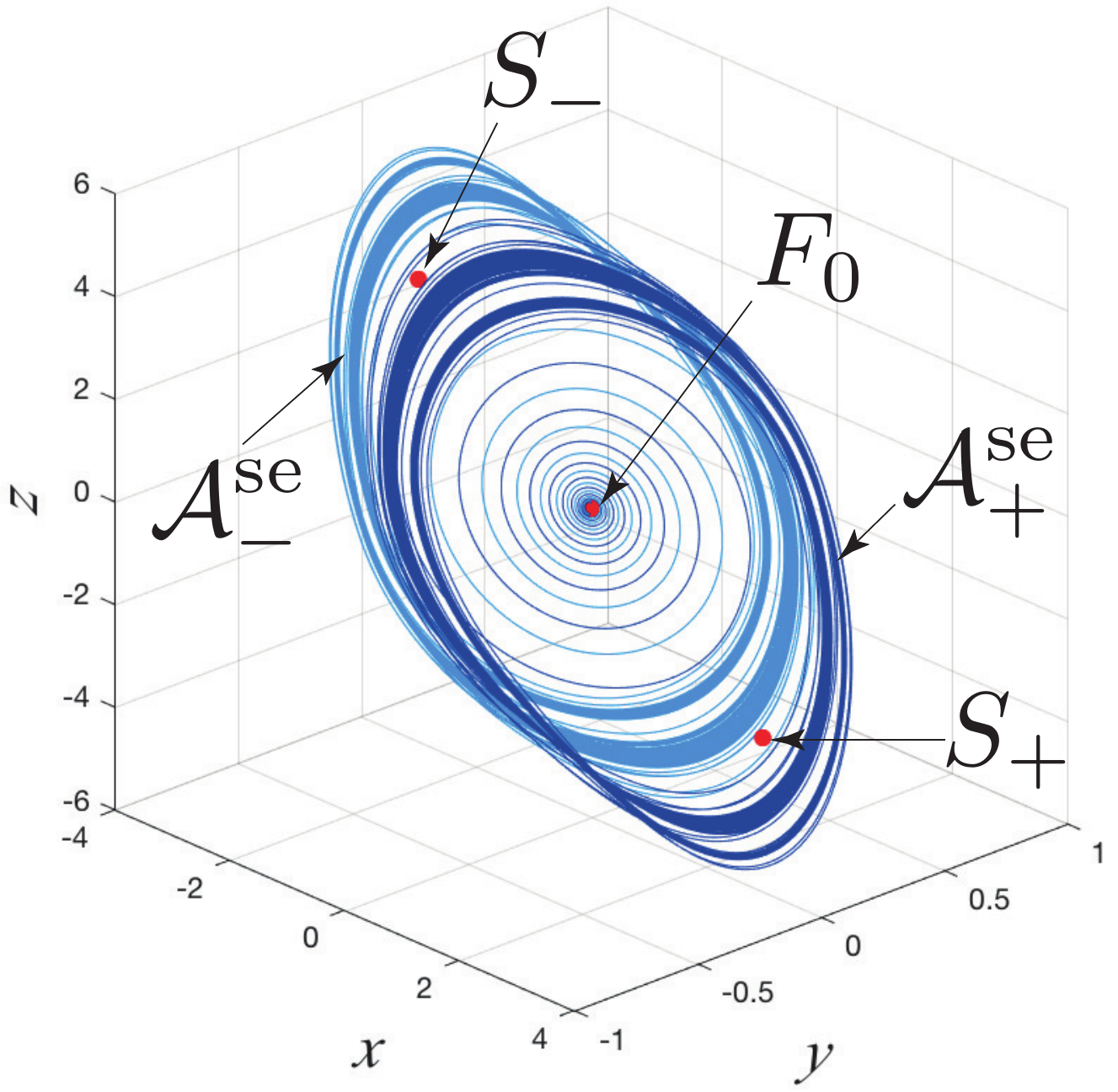}
  \includegraphics[width=0.45\textwidth]{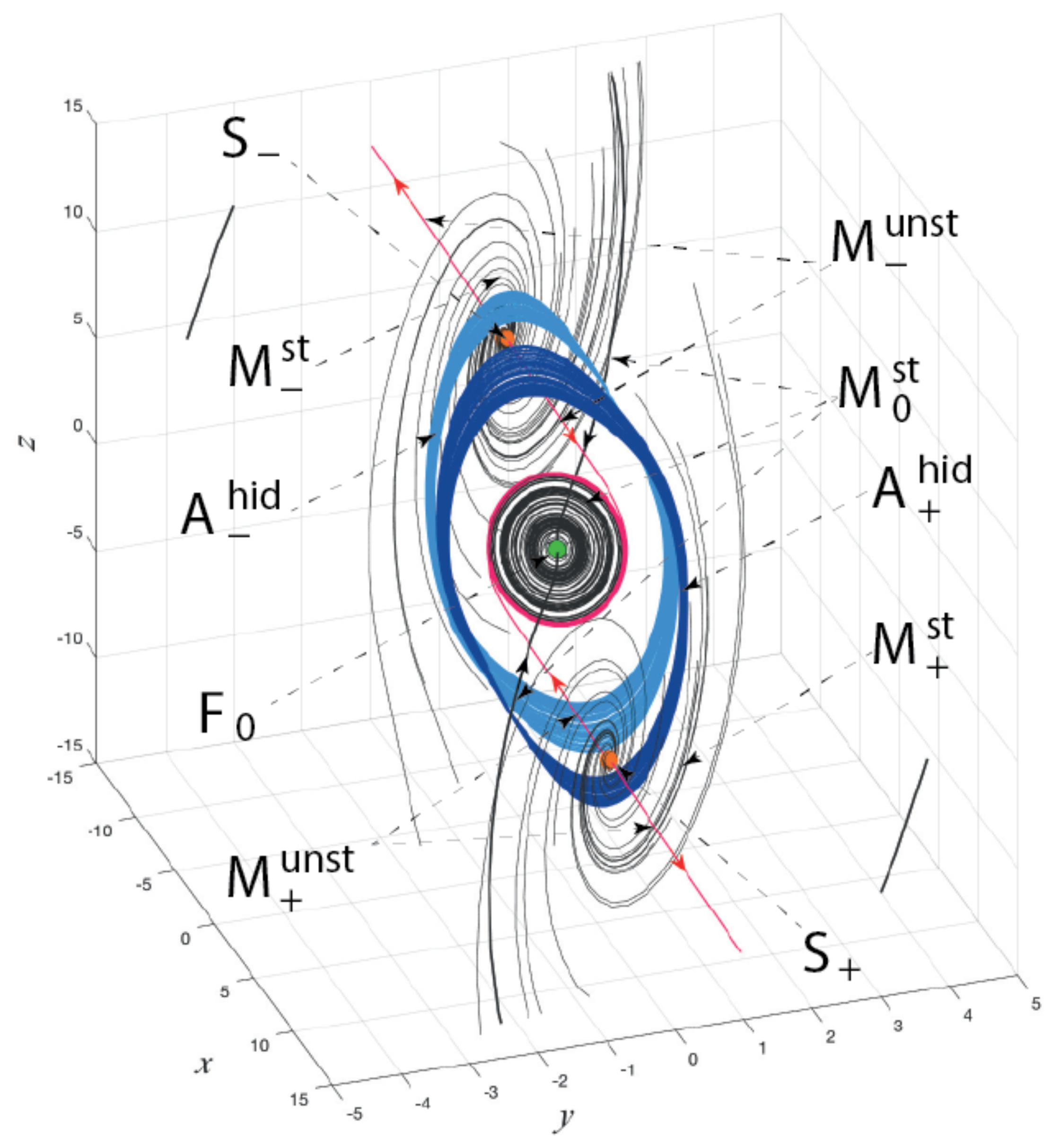}
  \caption{
  Left subfigure:
  two symmetric self-excited attractors (blue) excited from zero unstable equilibrium
  (parameters $\alpha = 15$, $\beta = 28$, $\gamma = 0$, $m_0 = -5/7$, $ m_1 = -8/7$).
  Right subfigure:
  two symmetric hidden chaotic attractors (blue),
  trajectories (red) from unstable manifolds ${\rm M_{\pm}^{\rm unst}}$
  of two saddle points $S_{\pm}$ are either attracted to
  locally stable zero equilibrium $F_0$, or tend to infinity;
  trajectories (black) from stable manifolds ${\rm M_{0,\pm}^{\rm st}}$
  tend to $F_0$ or $S_{\pm}$
  (parameters $\alpha = 8.4562$, $\beta = 12.0732$,
  $\gamma = 0.0052$, $m_0 = -0.1768$, $m_1 = -1.1468$).
  }
  \label{chua-se-ha-matlab}
\end{figure}

For example, hidden attractors are attractors in systems
without equilibria or with only one stable equilibrium
(a special case of multistability and coexistence of attractors).
The \emph{hidden vs self-excited classification of attractors}
was introduced
in connection with the discovery of the first hidden Chua attractor \cite{LeonovK-2009-PhysCon,KuznetsovLV-2010-IFAC,LeonovKV-2011-PLA,LeonovKV-2012-PhysD,KuznetsovKLV-2013}.
The \emph{Leonov-Kuznetsov's classification of attractors as hidden or self-excited}
is captured much attention of scientists from around the world
and hidden Chua attractors have become intensively studied
(see, e.g. \cite{LiZY-2014-HA,BurkinK-2014-HA,LiSprott-2014-HA,ZhusubaliyevMCM-2015-HA,KuznetsovKMS-2015-HA,ChenLYBXW-2015-HA,Semenov20151553,MenacerLC-2016-HA,Zelinka-2016-HA,DancaKC-2016,Danca-2016-HA,WeiPKW-2016-HA,PhamVJVK-2016-HA,JafariPGMK-2016-HA,DudkowskiJKKLP-2016}.
In Fig.~\ref{chua-se-ha-matlab} is shown an example of self-excited and hidden Chua
attractors visualized by numerical integration of system
\eqref{chua-dimensionless} in MATLAB.

\medskip

\section{Visualization of hidden Chua attractor in SPICE}

Nowadays various Simulation Programs with Integrated Circuit Emphasis (SPICE)
are widely used to analyze and design  analog circuits \cite{Williams-2016}.
Consider simulation of hidden Chua attractor in SIMetrix SPICE\footnote{
\url{https://www.simetrix.co.uk/}
}
for the following parameters:
$C_1 = 118.2u,\  C_2 = 1m,\ L = 82.8281,\ R0 = 430.7m,\ R = 1k,\ G_b = -0.0001768,\ G_a = -0.0011468$ ($\alpha = 8.4562$, $\beta = 12.0732, \gamma = 0.0052, m_0 =-0.1768, m_1 = -1.1468$).

\begin{lstlisting}[numbers=left, basicstyle=\small]
*#SIMETRIX
R1 L_P 0 430m
X$psi psi_OUT psi_OUT $$arbsourcepsi pinnames: N1 OUT
.subckt $$arbsourcepsi N1 OUT
B1 OUT 0 I=-0.0011468*V(N1) +  (-0.00017680+0.0011468)*LIMIT(V(N1),-1,1)
.ends
C1 psi_OUT 0 118.2u IC=2  BRANCH={IF(ANALYSIS=2,1,0)}
C2 C2_P 0 1m IC=1  BRANCH={IF(ANALYSIS=2,1,0)}
L L_P C2_P 82.8281 IC=-4m  BRANCH={IF(ANALYSIS=2,0,1)}
R psi_OUT C2_P 1k
.graph "XY(C2_P, psi_OUT)"  initxlims=false
.TRAN 0 50 0 10m UIC
\end{lstlisting}

\begin{figure}[h]
  \centering
  \includegraphics[width=0.8\textwidth]{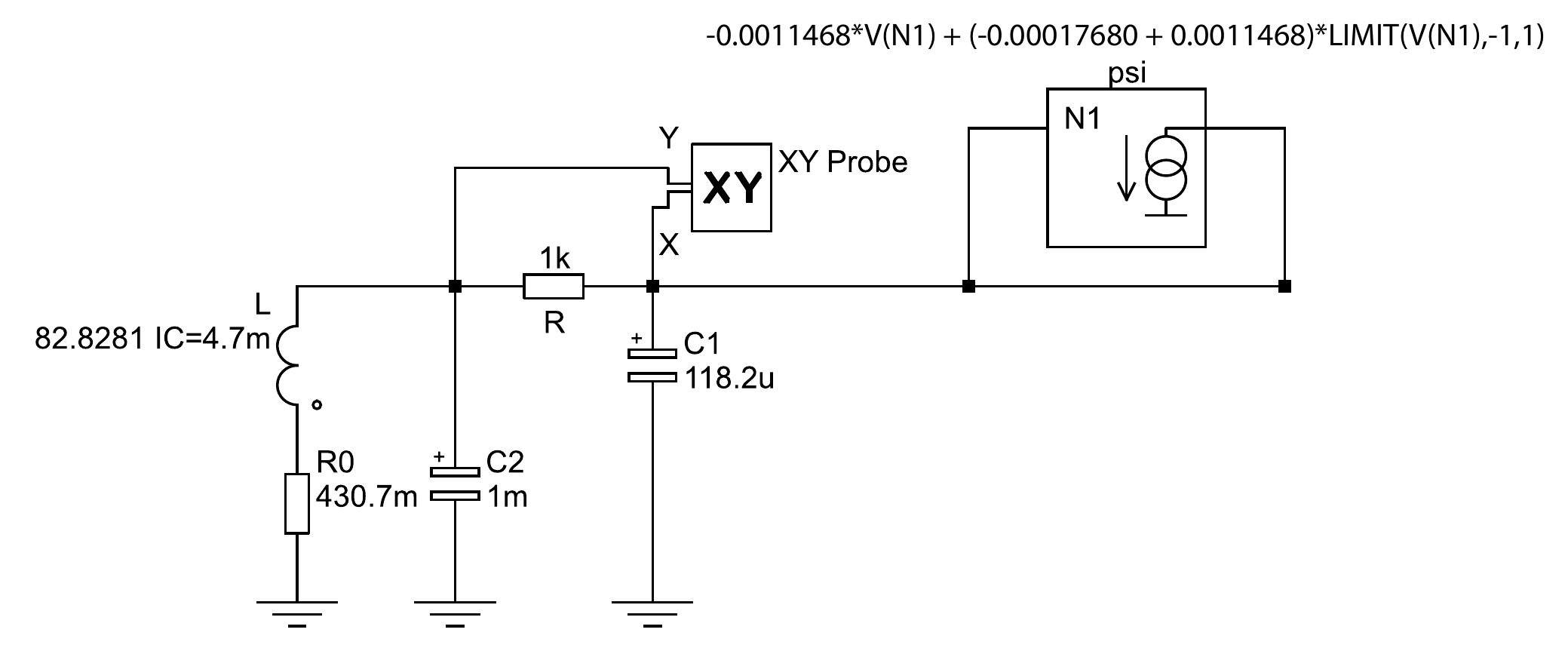}
  \caption{Chua circuit in SIMetrix SPICE}
  \label{chua-circuit-spice}
\end{figure}

In Fig.~\ref{chua-circuit-spice} is shown corresponding SPICE realization
of Chua circuit (see Fig.~\ref{chua-circuit-pic}).
Here $L, R0, R, C1, C2$ correspond to the elements of Chua circuit,
and element $XY Probe$ is used to measure the voltage on capacitors
(and plot projection of trajectories on $(v_1,v_2)$-plane).
The Chua diode with characteristic $f(v_1)$
is realized as ``Arbitrary Source'' (voltage-controlled current source) \textbf{psi}.

 \begin{figure}[h]
  \centering
  \includegraphics[width=0.8\textwidth]{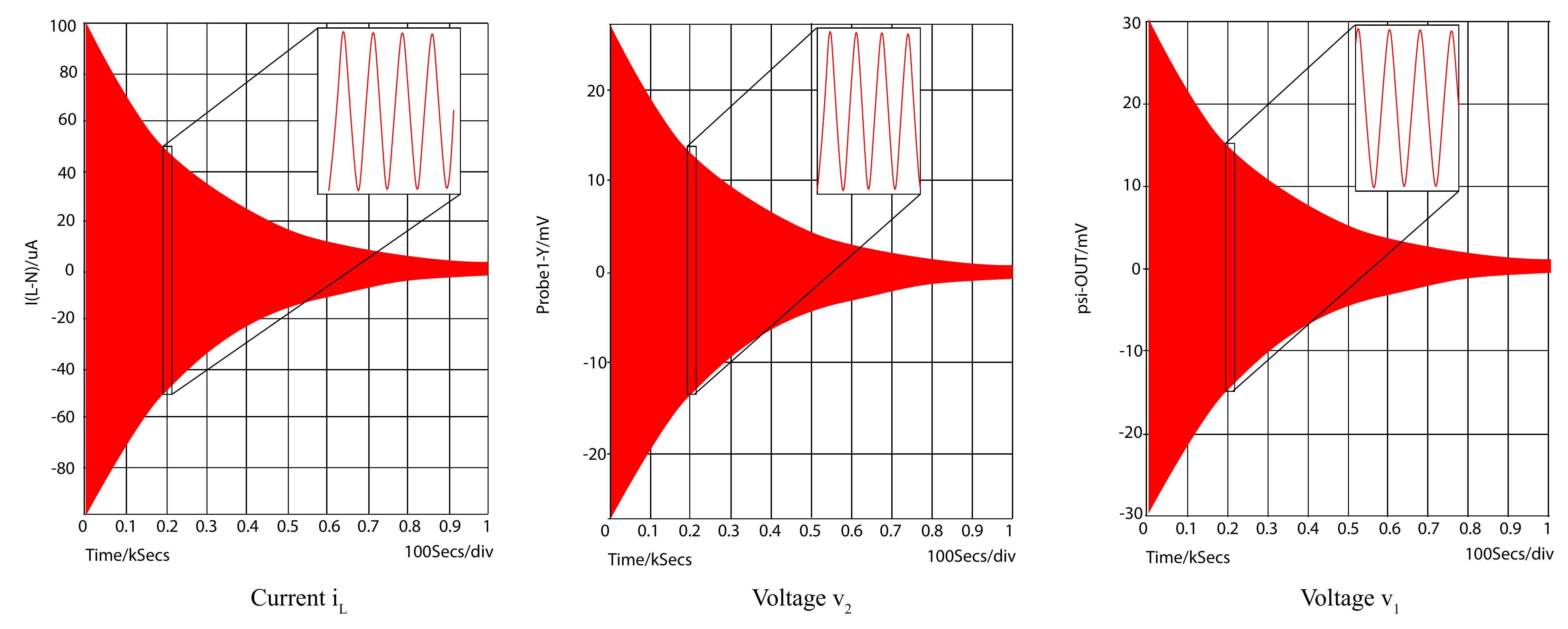}
  \caption{Vicinity of zero equilibrium: $v_2(0) = 0, \quad v_1(0) = 0,
  \quad i_L(0) = 100u$}
  \label{spice-result-100u1}
\end{figure}

\begin{figure}[h]
  \centering
  \includegraphics[width=0.8\textwidth]{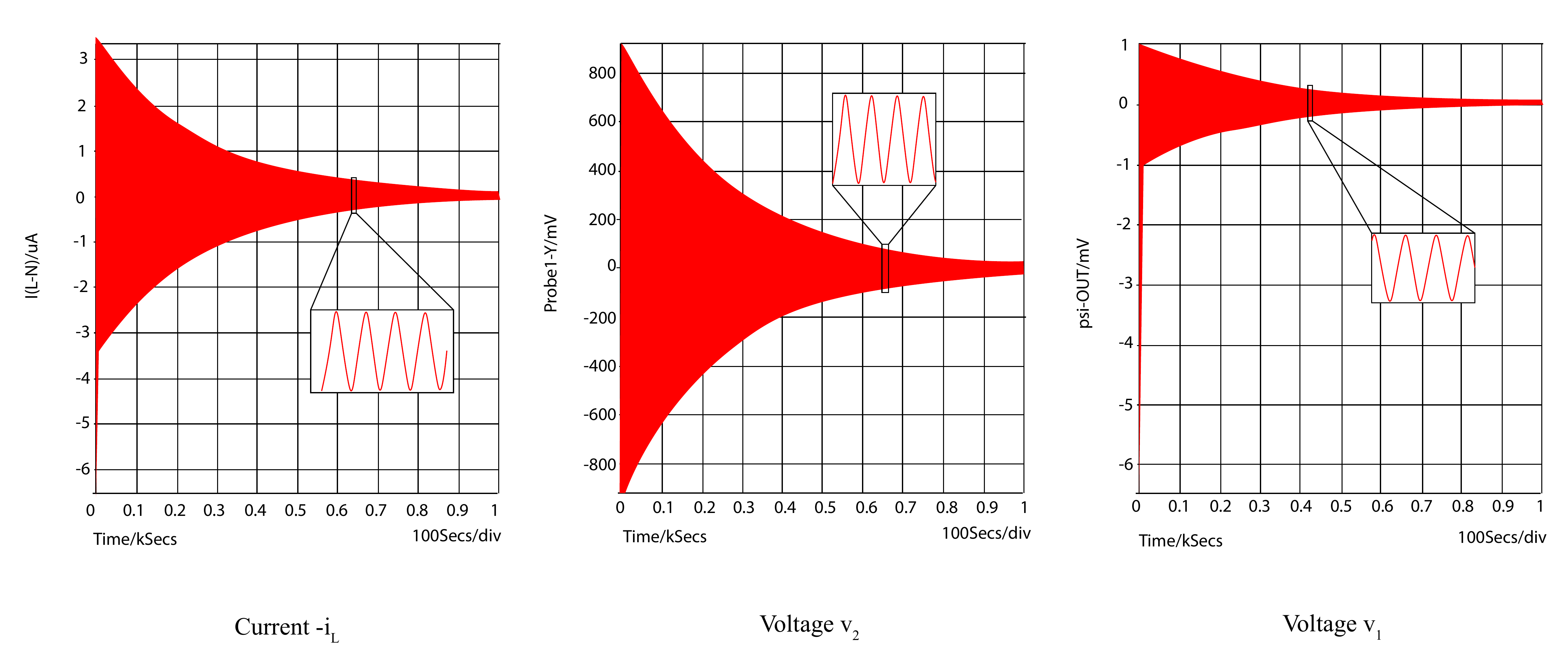}
  \caption{Vicinity of saddle point:
  $v_2(0) = -2.80804m, \quad v_1(0) = -6.5224, \quad i_L(0) = 6.5196m$}
  \label{spice-result-100u2}
\end{figure}

\begin{figure}[h]
  \centering
  \includegraphics[width=0.6\textwidth]{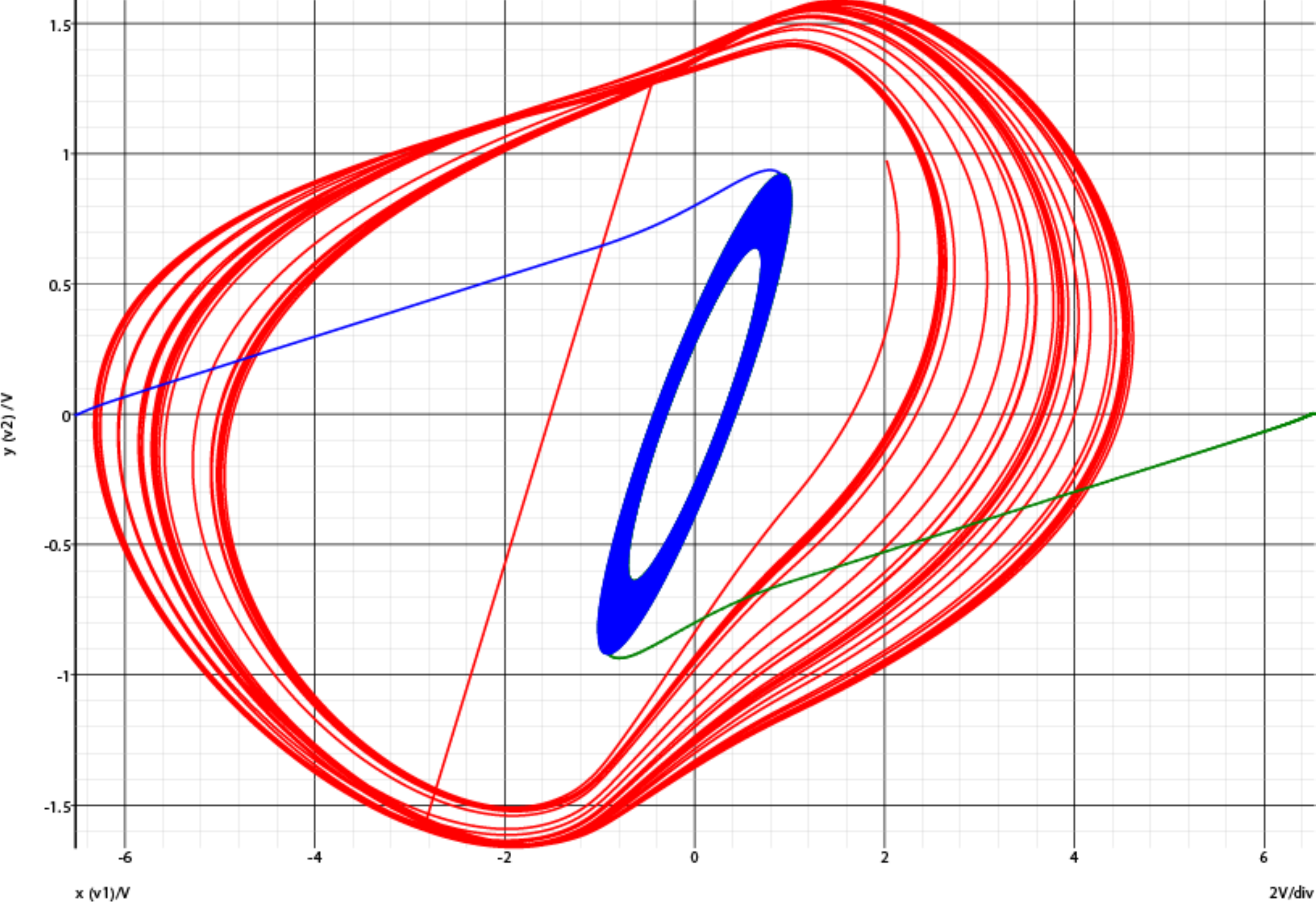}
  \caption{Voltage $v_1$ vs voltage $v_2$.
  Initial condition for blue trajectory (attracted to the stable zero equilibrium):
  $v_2(0) = -2.80804m, v_1(0) = -6.5224, i_L = 6.5196m$.
  Initial condition for red trajectory
  (visualizing a hidden attractor -- projection on the plane):
  $v_2(0) = 1, v_1(0) = 2, i_L = -4m$.
  Initial condition for green trajectory (attracted to the stable zero equilibrium): $v_2(0) = 2.80804m, v_1(0) = 6.5224, i_L = -6.5196m$.}
  \label{spice-result-1a}
\end{figure}

For the considered values of parameters there
are three equilibria in the system:
the zero equilibrium $F_0 = (0, \, 0, \, 0)$ is a stable focus-node
and two symmetric saddle equilibria $S_{\pm}$.
To check that an attractor is hidden,
we have to demonstrate that the trajectories from certain small vicinities
of equilibria are not attracted by the attractor.
Figs.~\ref{spice-result-100u1} and \ref{spice-result-100u2}
show SPICE simulation\footnote{
  In our experiment ``Max time step'' is set to $10m$.
}
of trajectory in a vicinity of zero equilibrium
and a trajectory with initial condition corresponding to the unstable manifold
of the saddle (in both cases the considered trajectories are attracted to the zero equilibrium).
Projection of twin symmetric chaotic attractors on the plane $(v_1,v_2)$
and attraction to the zero equilibrium are shown in Fig.~\ref{spice-result-1a}.

\begin{figure}[h]
  \centering
  \includegraphics[width=0.6\textwidth]{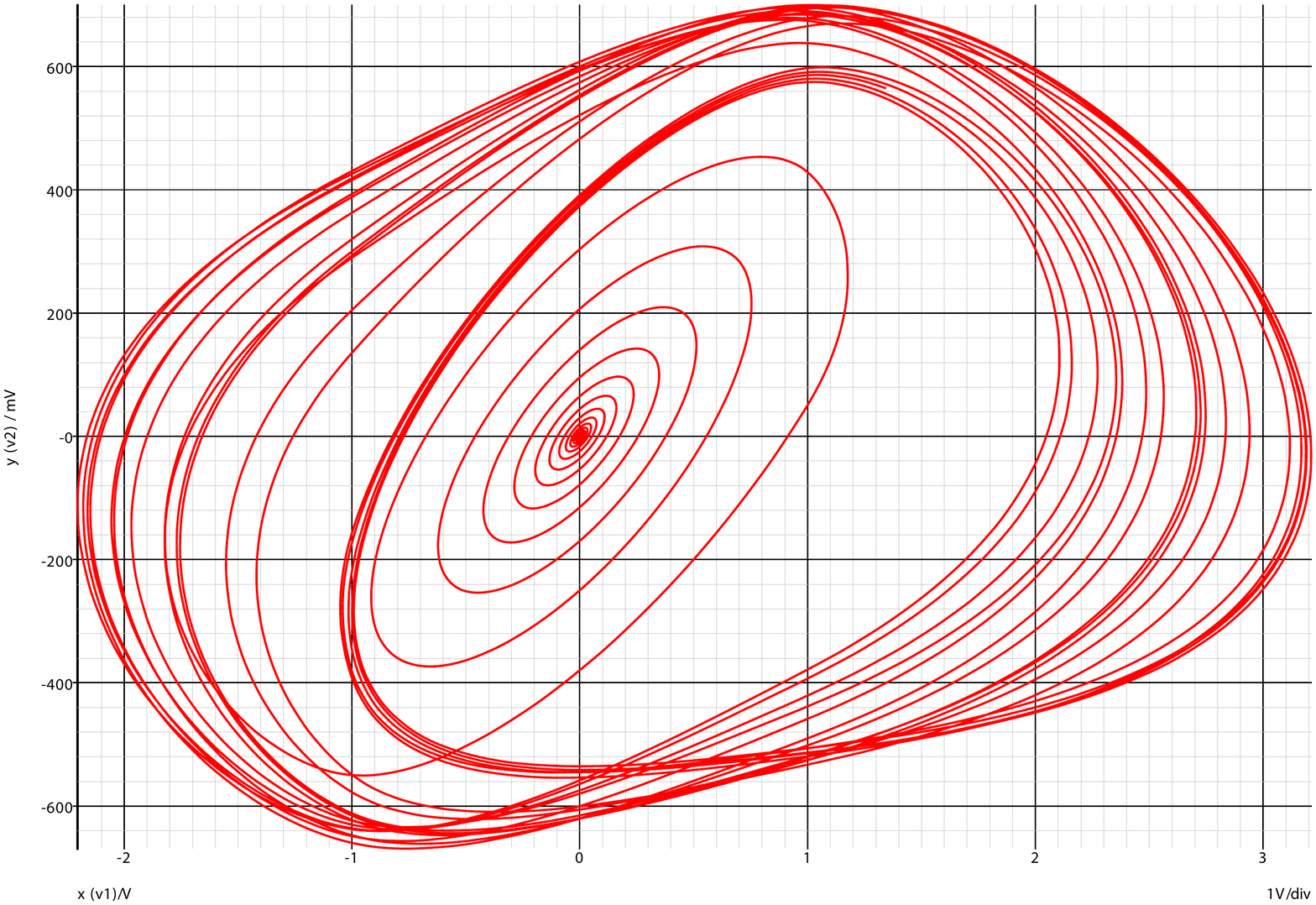}
  \caption{Voltage $v_1$ vs voltage $v_2$.
  Initial condition for red trajectory
  (visualizing a self-excited attractor – projection on the plane):
  $v_2(0) = 10m, \quad v_1(0) = 0, \quad i_L = 0$.
}
  \label{spice-result-1b}
\end{figure}

In Fig.~\ref{spice-result-1b} is shown simulation self-excited Chua attractor in SPICE
for the parameters
$C_1 = 66.667u,\  C_2 = 1m,\ L = 35.7143,\ R0 = 0,\ R = 1k,\ G_b = -0.00071429,\ G_a = -0.0011468$.
In this case the trajectory with initial data in a vicinity of zero
equilibrium is attracted by the self-excited attractor.

\section{Synchronization of Chua circuits}
Nowadays various chaotic secure communication systems based on
Chua circuits and other electronic generators of chaotic oscillations
are of interest
\cite{Kapitaniak-1992-chaotic,Ogorzalek-1997-chaos,Yang-2004,Eisencraft-2013-chaotic,KuznetsovL-2014-IFACWC,KuznetsovLMS-2016-INCAAM}.
The operation of such systems is based on the synchronization
chaotic signals of two chaotic identical generators (transmitter and receiver)
for different initial data.
The control signal proportional to the difference between the circuits signals,
adjust the state of one receiver.
The multistability and existence of hidden attractors
may lead to improper workreceiver of such systems.

\begin{figure}[H]
  \centering
  \includegraphics[scale=0.7]{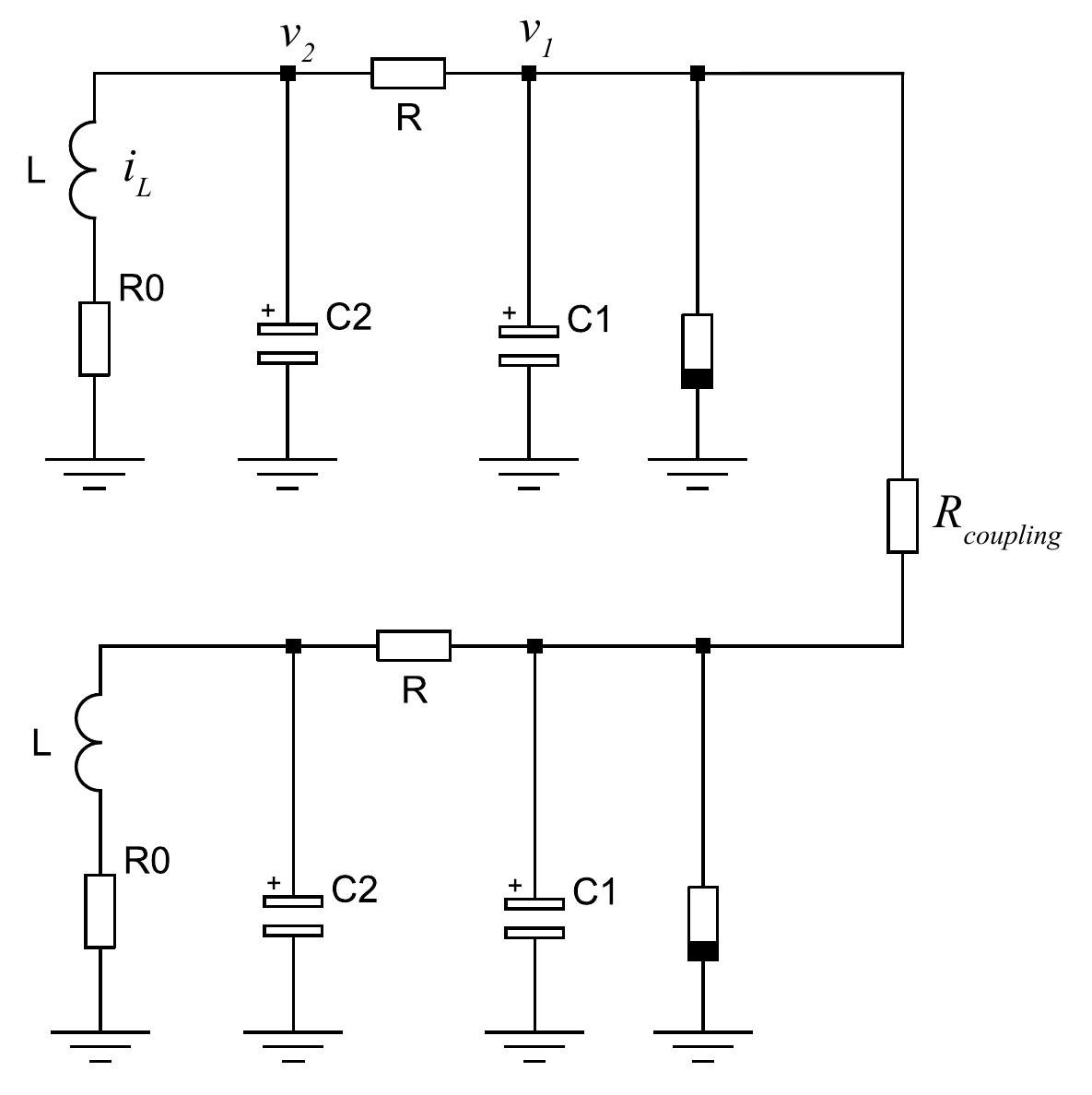}
  \caption{Two resistor-coupled Chua circuits}
  \label{chua-x-coupled-fig}
\end{figure}

Consider now two $x$-coupled Chua systems
\begin{equation}
\label{chua-x-coupled-dimensionless}
\begin{aligned}
& \frac{dx}{dt} = \alpha(y - x - \psi(x) + \delta(\tilde x - x)),
\\
& \frac{dy}{dt} =  x - y + z,
\\
& \frac{dz}{dt} = - (\beta y + \gamma z),
\\
& \frac{d\tilde x}{dt} = \alpha(\tilde y - \tilde x - \psi(\tilde x) + \delta(x - \tilde x)),
\\
& \frac{d\tilde y}{dt} =  \tilde x - \tilde y + \tilde z,
\\
& \frac{d\tilde z}{dt} = - (\beta \tilde y + \gamma \tilde z)б,
\end{aligned}
\end{equation}
where $\delta$ is a coupling factor ($\delta = \frac{R}{R_{\text{coupling}}}$),
and the corresponding circuit (see Fig.~\ref{chua-x-coupled-fig}).

\begin{figure}[htp]
  \centering
  \includegraphics[width=\textwidth]{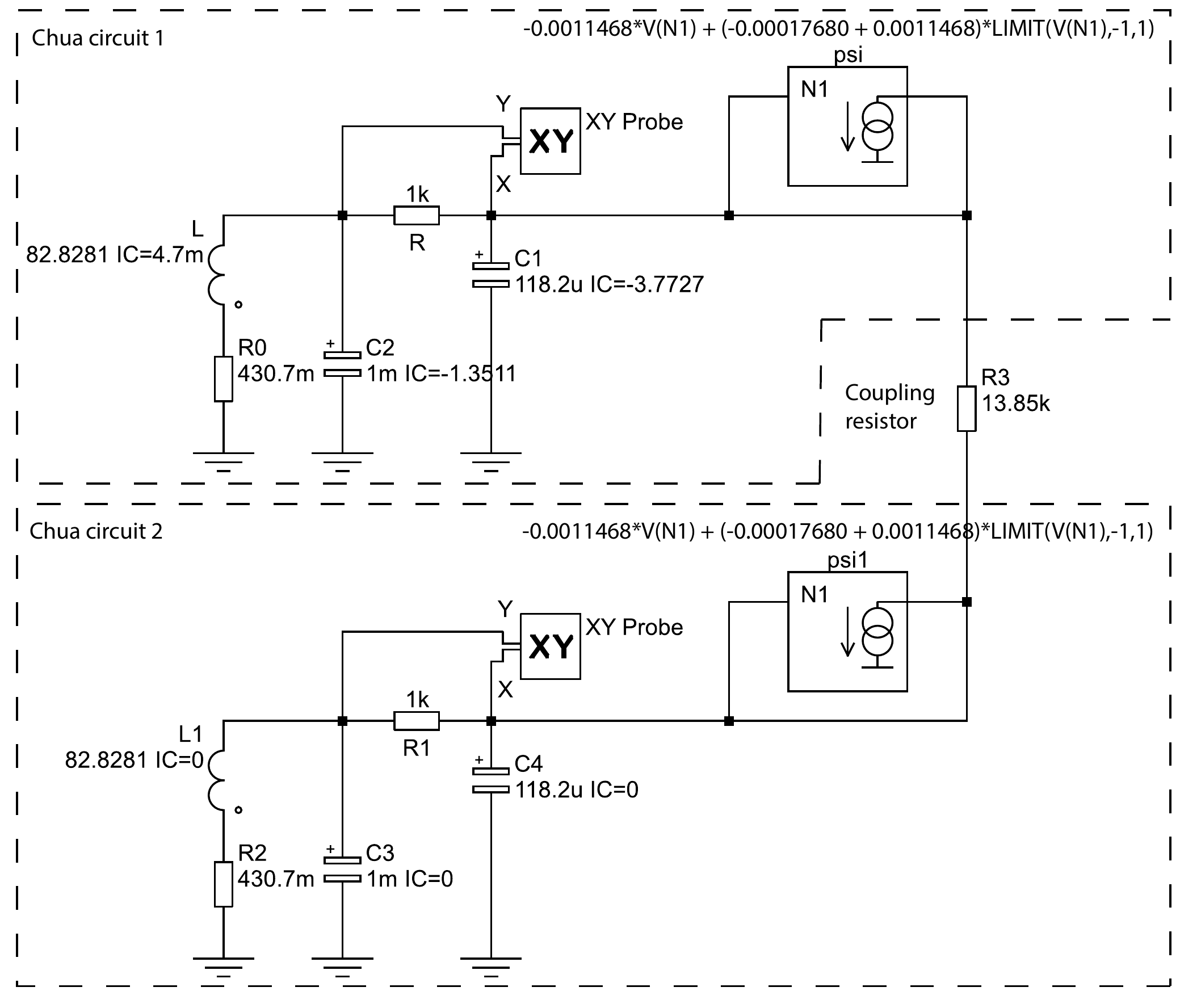}
  \caption{SIMetrix SPICE model of two coupled Chua's circuits}
  \label{x-coupled-chua-circuit-spice}
\end{figure}

SIMetrix SPICE realization of the coupled Chua circuits
is shown in Fig.~\ref{x-coupled-chua-circuit-spice}.
Here the passive elements of the first circuit: $L, R0, R, C1, C2$,
are identical to the corresponding elements of the second circuit: $L1, R1, R2, C3, C4$.
Elements $XY Probe$ are used to measure voltages on capacitors
and plot on $(v_1,v_2)$-plane.
To simulate nonlinear elements with characteristic $f(v_1)$,
we use SPICE elements ``Arbitrary Source'' (voltage-controlled current source)
\textbf{psi} and \textbf{psi1}.
Resistor $R3$ connects two circuits
and characterizes the distance between transmitter and receiver.
Below it is shown that critical coupling value
(i.e. maximum value for which synchronization takes place)
is different for different choice of initial data
because of the multistability and hidden attractors.
In our experiments the simulation time is $100$ seconds
and minimal value of $R3$ is 1000.


{\bf Example 1.}
 Consider initial data of the circuits on one of the symmetrical hidden attractors:
 Chua circuit 1 --- $v_1(0) = -3.7727, v_2(0) =  -1.3511, i_L(0) = 0.0047$;
 Chua circuit 2 --- $v_1(0) = -3.6, v_2(0) =  -1.3511, i_L(0) = 0.0047$.

 \begin{figure}[H]
  \centering
  \includegraphics[width=\textwidth]{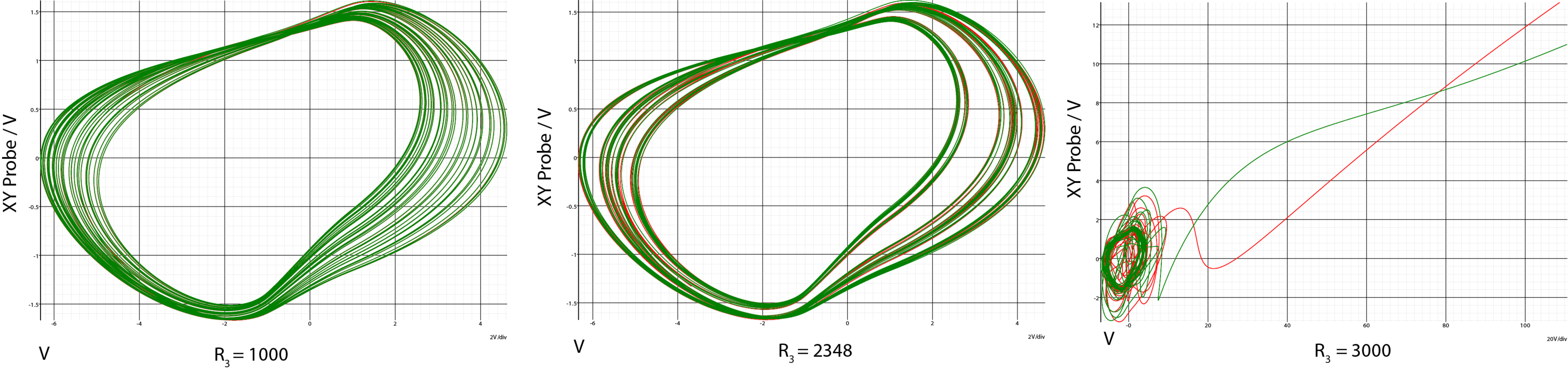}
  \caption{
  Initial data of the circuits on the same hidden attractor.
  Critical coupling value $R3 = 2.348 \cdot 10^3$.}
  \label{spice-result-2}
\end{figure}

{\bf  Example 2.}
Consider initial data of the circuits on two symmetrical hidden attractors:
Chua circuit 1: $v_1(0) = -3.7727, v_2(0) =  -1.3511, i_L(0) = 0.0047$;
Chua circuit 2: $v_1(0) = 3.7727, v_2(0) =  1.3511, i_L(0) = -0.0047$.

 \begin{figure}[h]
  \centering
  \includegraphics[width=\textwidth]{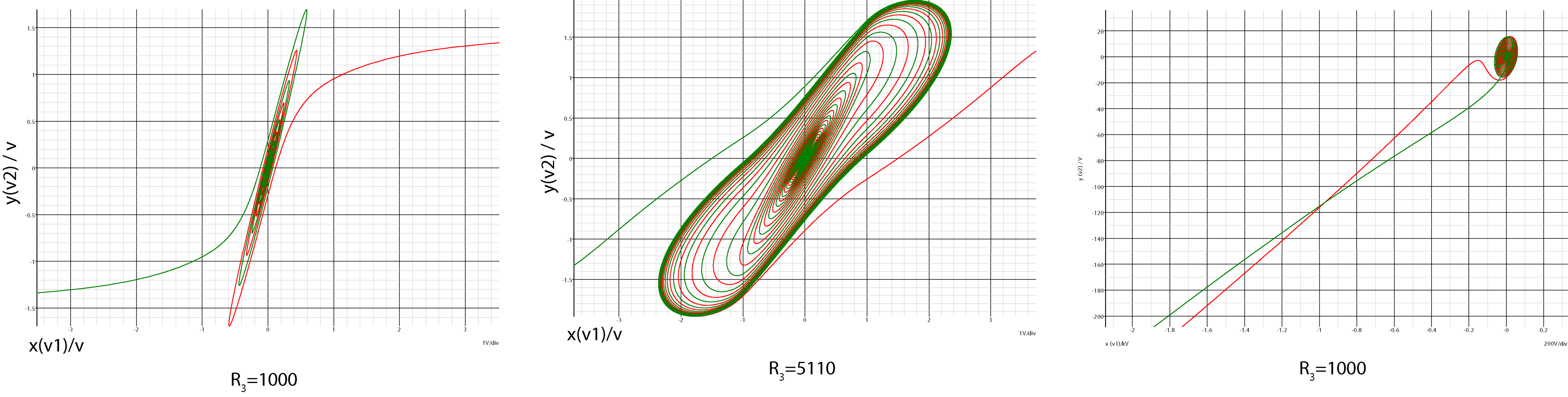}
  \caption{
  Initial data  of the circuits on two symmetrical hidden attractors.
  Critical coupling resistor value is $R3 = 5.110 \cdot 10^3$.
  }
  \label{spice-result-1}
\end{figure}

{\bf Example 3}
 Consider initial data of the circuits on one of the symmetrical hidden attractors
 and stable zero equilibrium:
 Chua circuit 1 --- $v_1(0) = -3.7727, v_2(0) =  -1.3511, i_L(0) = 0.0047$;
 Chua circuit 2  --- $v_1(0) = 0, v_2(0) =  0, i_L(0) = 0$.

 \begin{figure}[h]
  \centering
  \includegraphics[width=\textwidth]{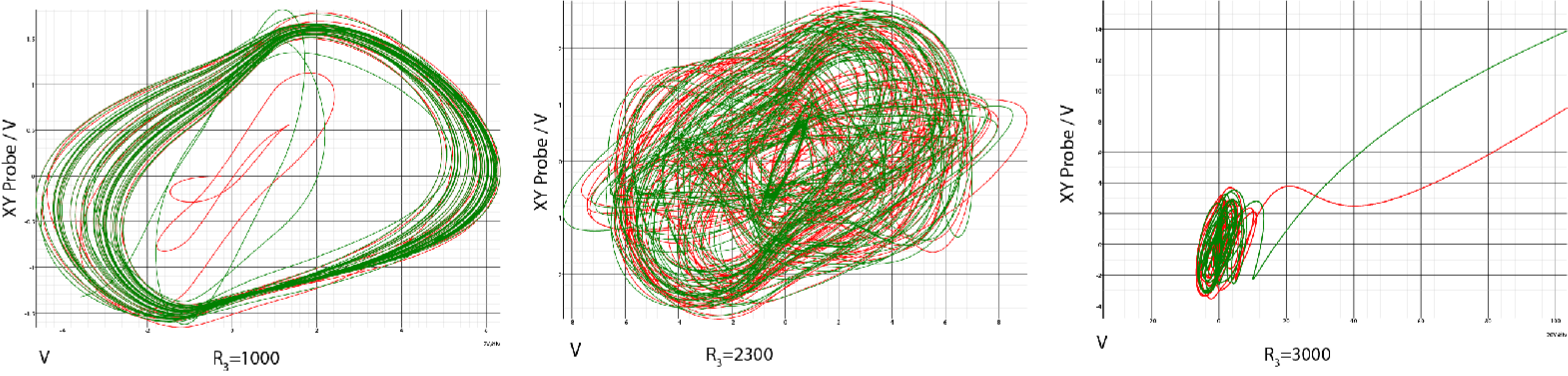}
  \caption{
  Initial data  of the circuits on a hidden attractor and stable zero equilibrium.
  Critical coupling resistor value is $R3 = 2.300 \cdot 10^3$.
  }
  \label{spice-result-2}
\end{figure}

{\bf Example 4}
 Consider the case of symmetrical self-excited attractors
 for the parameters $\alpha = 15$, $\beta = 28$, $\gamma = 0$, $m_0 = -5/7$, $ m_1 = -8/7$ (corresponding SPICE parameters: $C_1 = 6.667 \cdot 10^{-5},\  C_2 = 0.001,\  G_a = -7.1429 \cdot 10^{-4},\  G_b = -1.1429,\  L = 35.7134,\  R_0 = 0$).
 Take the following initial data of the circuits on
 one of the symmetrical self-excited attractors and unstable zero equilibrium:
 Chua circuit 1 --- $v_1(0) = -0.780, v_2(0) =  -0.525, i_L(0) = 3.5m$;
 Chua circuit 2  --- $v_1(0) = 0, v_2(0) =  0, i_L(0) = 0$.

 \begin{figure}[h]
  \centering
  \includegraphics[width=\textwidth]{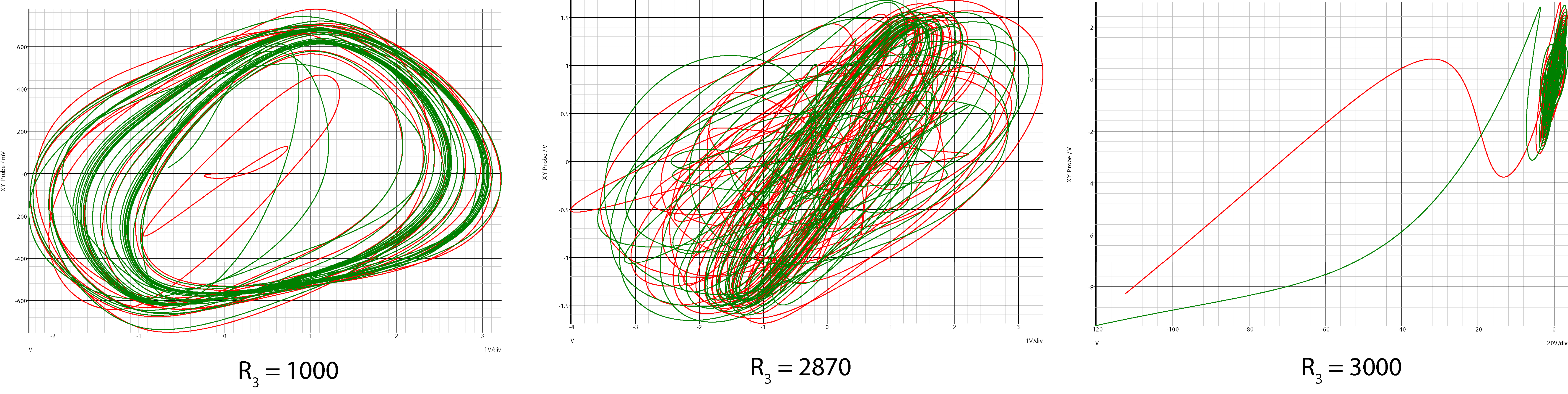}
  \caption{
  Initial data  of the circuits on a self-excited attractor and unstable zero equilibrium.
  Critical coupling resistor value is $R3 = 2.870 \cdot 10^3$.
  }
  \label{spice-result-3}
\end{figure}

\FloatBarrier

\section*{Conclusion}

 In conclusion we note that the simulation of circuit behavior by software,
as well as the numerical integration of its dynamical model,
is subject to numerical errors due to time discretization step
(see, e.g. the corresponding examples with phase-locked-loop based circuits in
\cite{LeonovKYY-2015-TCAS,BianchiKLYY-2016,BestKLYY-2016,KuznetsovLYY-2017-CNSNS}).
Observation of hidden Chua attractors in physical experiments
is discussed, e.g., in \cite{ChenLYBXW-2015-HA,BaoHCXY-2015-HA,ChenYB-2015-HA,ChenYB-2015-HA-EL,BaoLWX-2016}.

\bigskip
{\bf Acknowledgements.}

This work is supported by Leading Scientific School of Russian Federation (8580.2016.1, s.2)
and DST-RFBR project (16-51-45062, s.3).


\end{document}